\documentclass[aps,prc,twocolumn,superscriptaddress,floatfix,longbibliography]{revtex4-2}

\usepackage{graphicx}
\usepackage{amsmath,amssymb}
\usepackage{xcolor}
\usepackage[colorlinks=true,linkcolor=blue,citecolor=blue,urlcolor=blue]{hyperref}

\newcommand{\czero}{c_{0}}
\newcommand{\cone}{c_{1}}

\begin{document}

\title{Improved nonflow control for multiparticle correlators and system-size-dependent evolution from OO to Pb--Pb collisions at the LHC}

\author{Neelkamal Mallick}
\email[]{neelkamal.mallick@cern.ch}
\affiliation{Department of Physics, University of Jyv\"askyl\"a, P.O. Box 35, FI-40014, Jyv\"askyl\"a, Finland}

\author{Suraj Prasad}
\email[]{suraj.prasad@cern.ch}
\affiliation{HUN-REN Wigner Research Centre for Physics,
29--33 Konkoly-Thege Mikl\'os Str., H-1121 Budapest, Hungary}
\date{\today}

\begin{abstract}
Multiparticle correlations of event-plane angles, known as symmetry plane correlators (SPCs), and correlations of anisotropic flow amplitudes of different harmonics, known as symmetric cumulants (SCs), help constrain the
initial conditions and the transport properties of the system formed in heavy-ion collisions. Such multiparticle correlations are robust against
nonflow biases in heavy-ion collisions; however, in small systems the nonflow contribution is sizeable and can distort SPCs and SCs at low multiplicity, mimicking a flow-like signal. Evidence of nuclear-geometry-induced collectivity in light-ion collisions, such as OO
and Ne--Ne, has been established at the LHC. Measurements of SPCs and SCs in these systems, and their comparison to heavy-ion collisions, can therefore clarify the system-size dependence of these correlations and the strength of nonlinear mode coupling. In this work, an improved nonflow subtraction framework based on multiplicity reweighting ($\czero$) and
$v_1$-estimator-based ($\cone$) scaling is validated for selected SPCs and SCs using a controlled toy model and the \textsc{Pythia~8~Angantyr} model,
where collective flow is absent by construction. We find that $\cone$ provides the more reliable nonflow control and apply it to predict the SPCs and normalized SCs in OO collisions at $\sqrt{s_{\rm NN}}=5.36$~TeV
with the \textsc{ampt} string-melting model. The results are compared with published Pb--Pb measurements from ALICE and from \textsc{ampt} as a function of midrapidity charged-particle multiplicity
$\langle N_{\rm ch}\rangle(|\eta|<0.5)$, placing light- and heavy-ion systems on a common system-size axis. 
\end{abstract}

\maketitle


\section{Introduction}
\label{sec:intro}

Azimuthal anisotropy is a key signature of the strongly interacting matter created in ultrarelativistic heavy-ion collisions~\cite{Voloshin:2008dg,Heinz:2013th,Luzum:2011mm}.
In a noncentral collision, the nuclear overlap is anisotropic in the transverse plane, arising from the almond-shaped collision geometry or from the event-by-event fluctuations of nucleon (and sub-nucleonic) positions.
Hydrodynamic evolution converts this initial spatial anisotropy $\varepsilon_n$ into momentum-space anisotropies, characterized by the harmonic amplitudes $v_n$ and the associated symmetry planes $\psi_n$.
Correlations among the $v_n$ and among the $\psi_n$ then provide complementary constraints on the initial state and on the transport properties of the medium.

Symmetric cumulants (SCs) and their normalized form (NSCs) measure the correlations of event-by-event fluctuations of anisotropic flow harmonic amplitudes, as studied extensively in Pb--Pb collisions~\cite{ALICE:2016kpq,ALICE:2017kwu,ALICE:2021adw}. They isolate genuine correlations between \(v_n\) and \(v_m\) beyond trivial factorization and are sensitive to both initial-state fluctuations and the nonlinear hydrodynamic response~\cite{Bilandzic:2013kga}. In parallel, correlations among symmetry planes, known as symmetry plane correlators (SPCs), encode how event-plane angles of different-order harmonics align. Early measurements using the scalar-product method~\cite{ATLAS:2014ndd} were shown to be biased by simultaneous correlations among flow amplitudes~\cite{Bilandzic:2020csw,ALICE:2023wdn}. The Gaussian estimator (GE) removes that bias and enables a cleaner extraction of SPCs and has been studied in Pb--Pb collisions at \(\sqrt{s_{\mathrm{NN}}}=2.76\) and \(5.02\)~TeV, including higher-order combinations involving up to five planes by ALICE~\cite{ALICE:2023wdn,ALICE:2024fus}. Hydrodynamic models describe the overall trends but often overestimate correlators that are particularly sensitive to nonlinear mode coupling~\cite{ALICE:2023wdn,ALICE:2024fus}.

A natural next step is to extend SPC and SC measurements from heavy ions to light-ion systems. Collisions of $^{16}$O and $^{20}$Ne at the LHC bridge the multiplicity gap between small ($pp$ and $p$A) and large (Pb--Pb) systems. Recent measurements have already established finite elliptic and triangular flow, $v_2$ and $v_3$, and strong evidence of nuclear-geometry-driven collectivity in OO and Ne--Ne collisions~\cite{ALICE:2025luc,ATLAS:2025nnt,CMS:2025tga}.
Those single-harmonic observables, however, largely reflect the overall strength of the medium in converting initial-state spatial anisotropy and energy fluctuations into the final-state momentum space anisotropy.
The study of SPCs and SCs goes further as they probe the correlations among different symmetry planes and among different flow amplitudes. Together, these observables are therefore more sensitive than $v_n$ alone to nonlinear mode coupling and to finer details of the initial geometry, including features of the nuclear structure~\cite{Behera:2023nwj,Prasad:2026osg}. A system-size scan of SPCs and SCs can further reveal whether the same hydrodynamic response that describes Pb--Pb persists in light ions, and how initial-state geometry shapes multiparticle correlations from small to large systems~\cite{ALICE:2019zfl}.

In heavy-ion collisions, multiparticle correlators are typically dominated by long-range collective flow, while short-range few-body correlations---commonly called \emph{nonflow}---are comparatively small when the analysis is restricted to the soft sector (typically $p_{\rm T}\lesssim 3$~GeV/$c$),
where jet-driven contributions are suppressed~\cite{Voloshin:2008dg,Jia:2017hbm}. Nonflow arises from sources unrelated to the global collision geometry: jet fragmentation, resonance decays, quantum statistics (HBT), and other local particle clusters~\cite{Voloshin:2008dg,Wang:2026hak}. Although multiparticle cumulants and subevent methods can suppress nonflow contributions, residual contamination may remain, particularly at relatively low multiplicities, motivating further development and systematic validation of alternative approaches~\cite{Borghini:2001vi,Bilandzic:2010jr,Jia:2017hbm,ATLAS:2017rtr,Huo:2017nms,Nie:2018xog}.
Because these contributions involve only a limited number of correlated particles, their relative weight grows as the multiplicity $N$ decreases, scaling roughly as $1/N^{m-1}$ for an $m$-particle correlator in the independent-source picture~\cite{Borghini:2001vi,Wang:2026hak}. Consequently, in small and intermediate systems---where $N$ is much lower than in mid-central Pb--Pb---nonflow can dominate SPCs and SCs at low multiplicity and mimic a flow-like signal.
Reliable interpretation of these observables in OO and Ne--Ne therefore requires an explicit nonflow estimate and subtraction, beyond the partial suppression offered by rapidity gaps or subevent methods alone~\cite{Wang:2026hak}.


In this paper, improved nonflow subtraction methods are tested for SPCs and SCs in OO collisions at $\sqrt{s_{\rm NN}}~=~5.36$~TeV before being applied to the predictions from A Multi-Phase Transport (\textsc{ampt}) string-melting model.
Section~\ref{sec:method} introduces the observables and the nonflow subtraction framework based on multiplicity reweighting ($\czero$) and $v_1$-estimator scalings ($\cone$).
Section~\ref{sec:nfclosure} validates the procedure on a controlled toy model with known flow and nonflow, and in OO collisions at $\sqrt{s_{\rm NN}}=5.36$~TeV from \textsc{Pythia~8~Angantyr} model, where collective flow is absent by construction.
Section~\ref{sec:results} applies the prescription that best suppresses the nonflow to \textsc{ampt} string-melting OO collisions at the same energy and compares the results with published Pb--Pb measurements from ALICE and the \textsc{ampt} model. Section~\ref{sec:summary} summarizes the findings. Details of the toy-model generator, the \textsc{ampt} model, the scaling factors, and the remaining higher-order SPCs and NSCs are given in the appendices.

\section{Observables and nonflow subtraction}
\label{sec:method}

The nonflow subtraction methods used in this work are introduced in this section, together with the definitions of the prime observables, SPCs and normalized SCs. The choice of the nonflow estimator for each correlator and the domain of applicability are also mentioned.

\subsection{Low-multiplicity anchored subtraction}
In the independent-source picture, nonflow in an \(m\)-particle correlator comes from local clusters whose number grows with the multiplicity~\cite{Wang:2026hak}.
Of the \(\sim N^{m}\) distinct \(m\)-tuples in an event, only \(\sim N\) are drawn from one correlated source, so the nonflow contribution scales as
\begin{equation}
C^{\mathrm{nf}}
\propto \frac{1}{N^{m-1}}
\propto \langle v_1^{2}\rangle^{m-1}.
\label{eq:nfscale}
\end{equation}
The first relation is the multiplicity baseline (\(\czero\)).
The second is the dipole refinement (\(\cone\)): a correlator that contains \(v_1\) follows the same multiplicity dependence and absorbs the extra dependence induced by multiplicity selection.
Such correlators are dominated by nonflow, because the genuine dipolar flow changes sign with \(p_{\mathrm{T}}\) and nearly cancels after \(p_{\mathrm{T}}\) integration~\cite{Wang:2026hak}.

The nonflow template is the same correlator in the low-multiplicity (LM) class of the same data set, taken here as the \(70\)--\(100\%\) multiplicity class.
Collective flow in that class, if any, is neglected, which is justified when the target multiplicity lies well above the LM multiplicity~\cite{Wang:2026hak}.
The subtracted correlators are given as,\\
\textbf{$c_0$ scaling:}
\begin{align}
C^{\mathrm{sub}}
&= C^{\mathrm{obs}}
 - \left(\frac{N_{m,\mathrm{LM}}}{N_m}\right)^{m-1}
   C^{\mathrm{obs}}_{\mathrm{LM}}
\label{eq:c0}
\end{align}
\textbf{$c_1$ scaling:}
\begin{align}
C^{\mathrm{sub}}
&= C^{\mathrm{obs}}
 - \frac{E}{E_{\mathrm{LM}}}\,
   C^{\mathrm{obs}}_{\mathrm{LM}}.
\label{eq:c1}
\end{align}
Here, \(E\) is a nonflow estimator of the same particle order as the target and contains at least one power of \(v_1\).
The multiplicity in Eq.~\eqref{eq:c0} is fixed by the \(m\)-tuple average of \(1/N^{m-1}\)~\cite{Wang:2026hak},
\begin{align}
\left\langle \frac{1}{N^{m-1}}\right\rangle
&=
\frac{\sum_{i} M_m(N_i)\,N_i^{-(m-1)}}
{\sum_{i} M_m(N_i)}
\label{eq:Neff}\\
&= \frac{1}{N_m^{m-1}},
\nonumber
\end{align}
where the sum runs over events in the class and \(M_m(N)=N(N-1)\cdots(N-m+1)\) is the number of distinct \(m\)-tuples.
The same average in the LM class defines \(N_{m,\mathrm{LM}}\).
For \(N\gg m\), \(M_m(N)\approx N^{m}\) and \(N_m^{m-1}\) reduces to \(\sum_{i} N_i^{m}/\sum_{i} N_i\).
It satisfies \(N_m\ge\langle N\rangle\).
High-multiplicity events dominate that average inside a broad class, so \(N_m\) exceeds the mean multiplicity.
Using \(\langle N\rangle\) in place of \(N_m\) undersubtracts the nonflow, and the bias is largest in the wide LM class.

\paragraph*{Domain of applicability, assumptions, and advantages over other methods.}

The nonflow subtraction used in this work relies on the independent-source picture~\cite{Wang:2026hak}. Here, the \(1/N^{m-1}\) scaling of Eq.~\eqref{eq:nfscale} is the leading expectation and not an exact rule.
This scaling assumes that the measured correlator is a sum of genuine flow and nonflow, that the shape of the nonflow correlation does not change with multiplicity, and that the LM class has no genuine collective flow. It is important to note that any flow left in the LM template is subtracted along with the nonflow, which biases the final result~\cite{Wang:2026hak}.
Thus, the method is applicable only when the multiplicity of the target class is much larger than that of the LM class, i.e., \(N\gg N_{\mathrm{LM}}\), and at low \(p_{\mathrm{T}}\) (\(\lesssim 2\)~GeV/\(c\)), where the nonflow is dominated by dipolar correlations. At high \(p_{\mathrm{T}}\), or when a single jet dominates the event, the \(1/N^{m-1}\) scaling breaks down~\cite{Wang:2026hak}.
In this study, the observables are integrated over \(0.2<p_{\mathrm{T}}<5.0\)~GeV/\(c\), which goes beyond this \(p_{\mathrm{T}}\) limit. However, the closure test with \textsc{Pythia~8 Angantyr} in the same \(p_{\mathrm{T}}\) range shows that the subtraction removes most of the nonflow, leaving a residual at the \(10\%\) level or below for most of the observables (Table~\ref{tab:pythiaResiduals}). Moreover, the results are shown for the \(0\)--\(60\%\) classes, which are well separated from the \(70\)--\(100\%\) LM class.

The \(\cone\) scaling further assumes that the \(p_{\mathrm{T}}\)-integrated dipolar flow is negligibly small. The $v_1$-based estimator can therefore be used to test the assumed $1/N^{m-1}$ scaling of nonflow. It also accounts for deviations associated with multiplicity selection, which are not captured by the pure multiplicity scaling of $\czero$~\cite{Wang:2026hak}.
Moreover, for the SPCs, only the numerator is corrected, which assumes that the nonflow in the denominator \(D\) is small.
Reference~\cite{Wang:2026hak} has validated this framework only with the Heavy Ion Jet Interaction Generator (\textsc{hijing}), which has no flow, for \(\langle v_2^2\rangle\), \(\langle v_2^2\delta p_{\mathrm{T}}\rangle\), and \(c_2\{4\}\), and found that the best estimator differs from one observable to another. Therefore, before applying it to the SPCs and normalized SCs involving higher harmonics, we validate the method in Sec.~\ref{sec:nfclosure} using \textsc{Pythia~8} and a toy model, where the latter also tests whether a known flow signal is recovered.
In contrast to the subevent and pseudorapidity-gap methods, the template subtraction uses the full detector acceptance, thus retaining the statistical precision. Further, it can be applied in the same way to both SPCs and SCs using either LM class of the same data set or minimum-bias $pp$ collisions as the reference.


\subsection{Symmetry plane correlators}

\begin{table}[!ht]
\caption{Estimators \(E\) for the \(\cone\) subtraction of each symmetry plane correlator and normalized symmetric cumulant.
The \(\czero\) subtraction uses the same numerators with the multiplicity factor of Eq.~\eqref{eq:c0}.
Here, \(C_{112}=\langle V_{1}V_{1}V_{2}^{*}\rangle\), \(C_{123}=\langle V_{1}V_{2}V_{3}^{*}\rangle\), and \(c_{1}\{4\}=\langle v_{1}^{4}\rangle-2\langle v_{1}^{2}\rangle^{2}\).}
\label{tab:spcE}
\begin{ruledtabular}
\begin{tabular}{lll}
Observable & Numerator \(C\) & Estimator \(E\) \\
\colrule
\multicolumn{3}{l}{Symmetry plane correlators (SPC)} \\
\(\langle\cos(4\psi_{4}-4\psi_{2})\rangle\)
  & \(C_{422}=\langle V_{2}V_{2}V_{4}^{*}\rangle\)
  & \(C_{112}\) \\
\(\langle\cos(6\psi_{6}-6\psi_{3})\rangle\)
  & \(C_{633}=\langle V_{3}V_{3}V_{6}^{*}\rangle\)
  & \(C_{123}\) \\
\(\langle\cos(2\psi_{2}+3\psi_{3}-5\psi_{5})\rangle\)
  & \(C_{235}=\langle V_{2}V_{3}V_{5}^{*}\rangle\)
  & \(C_{123}\) \\
\(\langle\cos(2\psi_{2}+4\psi_{4}-6\psi_{6})\rangle\)
  & \(C_{246}=\langle V_{2}V_{4}V_{6}^{*}\rangle\)
  & \(C_{123}\) \\
\(\langle\cos(6\psi_{6}-6\psi_{2})\rangle\)
  & \(C_{6222}=\langle V_{2}V_{2}V_{2}V_{6}^{*}\rangle\)
  & \(c_{1}\{4\}\) \\
\colrule
\multicolumn{3}{l}{Normalized symmetric cumulants (NSC)} \\
NSC$(4,2)$ & SC$(4,2)$ & SC$(1,2)$ \\
NSC$(3,2)$ & SC$(3,2)$ & SC$(1,2)$ \\
NSC$(4,3)$ & SC$(4,3)$ & SC$(1,3)$ \\
NSC$(5,2)$ & SC$(5,2)$ & SC$(1,2)$ \\
NSC$(5,3)$ & SC$(5,3)$ & SC$(1,3)$ \\
\end{tabular}
\end{ruledtabular}
\end{table}

The \(m\)-particle azimuthal correlator is given as
\begin{equation}
\langle m\rangle_{n_1,\ldots,n_m}
=
\bigl\langle
e^{i(n_1\varphi_1+\cdots+n_m\varphi_m)}
\bigr\rangle,
\label{eq:mcorr}
\end{equation}
where $\varphi_i$ is the azimuthal angle of the $i$-th particle, and the average runs over all distinct sets of \(m\) particles in one event and then over the event sample~\cite{Bilandzic:2013kga,ALICE:2024fus}.
The integers \(a_i\) are the signed coefficients of the positive harmonics \(n_i\).
They satisfy \(\sum_i a_i n_i=0\), and the particle order is \(m=\sum_i|a_i|\).
Equation~\eqref{eq:mcorr} factorizes into the flow vectors \(V_n=v_n e^{in\psi_n}\), as
\begin{equation}
\begin{aligned}
\langle m\rangle_{n_1,\ldots,n_m}
&=
\bigl\langle V_{n_1}^{a_1}\cdots V_{n_k}^{a_k}\bigr\rangle\\
&=
\Bigl\langle
v_{n_1}^{|a_1|}\cdots v_{n_k}^{|a_k|}
\,e^{i(a_1 n_1\psi_{n_1}+\cdots+a_k n_k\psi_{n_k})}
\Bigr\rangle,
\end{aligned}
\label{eq:mflow}
\end{equation}
where $\psi_n$ is the $n$-th order symmetry plane angle. The index \(k\) counts only the distinct harmonics, so \(k\le m\).

The numerator used in the symmetry plane correlator is the real part of the measured correlator in Eq.~\eqref{eq:mflow},
\begin{equation}
\begin{aligned}
C 
&= \mathrm{Re}\,\langle m\rangle_{n_1,\ldots,n_m} \\ 
&=
\Bigl\langle
v_{n_1}^{|a_1|}\cdots v_{n_k}^{|a_k|}
\cos(a_1 n_1\psi_{n_1}+\cdots+a_k n_k\psi_{n_k})
\Bigr\rangle.
\end{aligned}
\label{eq:Cnum}
\end{equation}
The denominator is the \(2m\)-particle correlator in which every harmonic is paired with its complex conjugate, such that the event-plane angles are dropped,
\begin{equation}
\begin{aligned}
D
&=
\langle 2m\rangle_{n_1,-n_1,\ldots,n_m,-n_m}
=
\bigl\langle v_{n_1}^{2|a_1|}\cdots v_{n_k}^{2|a_k|}\bigr\rangle.
\end{aligned}
\label{eq:Dden}
\end{equation}
Using the Gaussian estimator (GE), the SPC is estimated by~\cite{Bilandzic:2020csw,ALICE:2024fus}
\begin{equation}
\bigl\langle
\cos(a_1 n_1\psi_{n_1}+\cdots+a_k n_k\psi_{n_k})
\bigr\rangle_{\mathrm{GE}}
=
\sqrt{\frac{\pi}{4}}\,\frac{C}{\sqrt{D}}.
\label{eq:ge}
\end{equation}
Because the magnitudes remain correlated inside \(D\), this estimator avoids the bias of the scalar-product form, whose denominator factorizes into single-harmonic moments assuming that event-by-event fluctuations of flow amplitudes are mutually independent. Both \(\langle m\rangle\) and \(\langle 2m\rangle\) are obtained from the recursive algorithm of the generic framework~\cite{Bilandzic:2013kga}.
The recursion builds each correlator from \(Q\)-vectors in one pass over the event and removes autocorrelations by construction. This also reduces the computational cost significantly. The measured numerator in Eq.~\eqref{eq:Cnum} still contains nonflow. That contribution is subtracted with Eqs.~\eqref{eq:c0} and~\eqref{eq:c1}; the denominator \(D\) is left unchanged, and the estimator in Eq.~\eqref{eq:ge} is rebuilt from the subtracted numerator. In the \(\cone\) scaling, the nonflow estimator $E$ is an estimator of the same particle order as the numerator itself: \(C_{112}\) and \(C_{123}\) for the three-particle correlators such as \(C_{422}\), \(C_{633}\), \(C_{235}\) and \(C_{246}\), and \(c_1\{4\}\) for the four-particle correlators such as \(C_{6222}\) (Table~\ref{tab:spcE}).


\subsection{Normalized symmetric cumulants}
The symmetric cumulant of two flow harmonics $m$ and $n$ is defined as a four-particle cumulant~\cite{Bilandzic:2013kga}
\begin{align}
\mathrm{SC}(m,n)
&= \langle 4\rangle_{m,n,-m,-n}
   - \langle 2\rangle_{m,-m}\,\langle 2\rangle_{n,-n} \nonumber\\
&= \langle v_m^2 v_n^2\rangle
   - \langle v_m^2\rangle\langle v_n^2\rangle,
\label{eq:sc}
\end{align}
where $\langle 2\rangle$ and $\langle 4\rangle$ are the two- and four-particle correlators of Eq.~\eqref{eq:mcorr}, with events weighted by the number of distinct pairs and quadruplets, respectively. In the absence of nonflow contributions, the SC measures the correlation between the event-by-event fluctuations of $v_m$ and $v_n$ through the covariance term of  $v_m^2$ and $v_n^2$ in Eq.~\eqref{eq:sc}. A positive (negative) value of SC$(m,n)$ indicates that the two harmonics are correlated (anticorrelated).

To reduce the dependence on the magnitudes of the flow harmonics, the SC is normalized as~\cite{ALICE:2016kpq,ALICE:2017kwu}
\begin{equation}
\mathrm{NSC}(m,n)
= \frac{\mathrm{SC}(m,n)}{c_m\{2\}\,c_n\{2\}},
\qquad
c_n\{2\}
= \langle 2\rangle_{n,-n},
\label{eq:nsc}
\end{equation}
where $c_n\{2\}$ estimates $\langle v_n^2\rangle$. All correlators are calculated using the $Q$-vector formalism~\cite{Bilandzic:2010jr,Bilandzic:2013kga}. The calculations are performed for the harmonic pairs $(m,n)=(3,2)$, $(4,2)$, $(4,3)$, $(5,2)$, and $(5,3)$. Results for NSC$(4,2)$ and NSC$(3,2)$ are presented in the main text, while those for the remaining combinations are given in Appendix~\ref{app:pythiaRest}.

Since the two terms of Eq.~\eqref{eq:sc} carry different event weights, averaging over a broad multiplicity class can produce a spurious SC even without genuine four-particle correlations. Following Ref.~\cite{Wang:2026hak}, SC$(m,n)$ is therefore calculated in unit-$N$ bins and averaged with the four-particle weights
\begin{equation}
W_4(N)=\sum_{\mathrm{events}} M_4(N),
\end{equation}
where $M_4(N)=N(N-1)(N-2)(N-3)$ is the number of distinct quadruplets, as in Eq.~\eqref{eq:Neff}.
The two-particle cumulants $c_n\{2\}$ used to calculate NSC$(m,n)$ are evaluated over the full multiplicity class.

\subsubsection{Nonflow subtraction using the template method}

Equations~\eqref{eq:c0} and~\eqref{eq:c1} are applied to the four-particle cumulant, $C=\mathrm{SC}(m,n)$. For the $\czero$ scaling,
\begin{equation}
\mathrm{SC}^{\mathrm{sub}}(m,n)
=\mathrm{SC}(m,n)-\left(\frac{N_{4,\mathrm{LM}}}{N_4}\right)^{3}\mathrm{SC}_{\mathrm{LM}}(m,n),
\end{equation}
with the four-particle effective multiplicities from Eq.~\eqref{eq:Neff}.
For the $\cone$ scaling, the estimator is the symmetric cumulant between the first harmonic and the lower harmonic $n$ of the target, $E=\mathrm{SC}(1,n)$,
\begin{equation}
\mathrm{SC}^{\mathrm{sub}}(m,n)
=\mathrm{SC}(m,n)-\frac{\mathrm{SC}(1,n)}{\mathrm{SC}_{\mathrm{LM}}(1,n)}\,\mathrm{SC}_{\mathrm{LM}}(m,n).
\end{equation}
Thus, SC$(1,2)$ is used for SC$(3,2)$, SC$(4,2)$ and SC$(5,2)$, while SC$(1,3)$ is used for SC$(4,3)$ and SC$(5,3)$ (Table~\ref{tab:spcE}). $\mathrm{NSC}^{\mathrm{sub}}(m,n)$ is obtained by replacing SC$(m,n)$ with $\mathrm{SC}^{\mathrm{sub}}(m,n)$ in Eq.~\eqref{eq:nsc}.

\subsection{Kinematics and multiplicity selection}
\label{subsec:kinematics}

In this study, the SPCs and NSCs are calculated using primary charged particles~\cite{ALICE:2017hcy} with $0.2~<~p_{\rm T}~<~5.0$~GeV/$c$ at midrapidity, $|\eta|<0.8$, similar to the acceptance of the ALICE detector at the LHC. For the multiplicity percentile estimation, the acceptance regions of the forward-timing (FT0) detectors of ALICE in Run~3~\cite{ALICE:2023udb}, known as the FT0M multiplicity, which is the sum of the amplitudes measured with FT0A ($3.5<\eta<4.9$) and FT0C ($-3.3<\eta<-2.1$), are used. This selection ensures a minimal autocorrelation among the tracks, since the tracks used for observables at midrapidity are different from the ones used for the estimation of the event multiplicity at forward rapidity. The multiplicity percentiles are then divided into nine classes with edges at 0, 5, 10, 20, 30, 40, 50, 60, 70, and 100\%, where the LM template belongs to the 70--100\% multiplicity class.

\section{Nonflow closure}
\label{sec:nfclosure}

In this section, the nonflow subtraction closure is demonstrated first on a toy model and then on \textsc{Pythia~8} to ensure the effectiveness of the methods in removing the nonflow as well as to estimate the nonflow residuals for each observable considered in this study. 

\subsection{Toy-model closure}
\label{subsec:toymodel}

The nonflow subtraction methods described in Sec.~\ref{sec:method} are validated on event samples generated using a toy model. The parameters of the toy model are presented in Appendix~\ref{app:toy}. This closure is crucial since the flow and nonflow pieces in the toy model are known separately. Three samples with an equal number of events are generated: one with only flow, one with only nonflow, and one with both (mixed). All three samples use the nine multiplicity classes as stated in Sec.~\ref{subsec:kinematics}. 

In each event a fraction \(f\simeq0.18\) of the tracks, or at least six, is reserved for nonflow clusters, and the remaining tracks form the bulk. This fraction is held fixed so that the nonflow contribution stays finite at high multiplicity. The same split is used in every sample, so at a given multiplicity the three samples have the same number of tracks in the reserved groups and the same number in the bulk. This is called \textit{occupancy-matching}. Each such reserved group of tracks forms a cluster of six tracks in the nonflow and mixed samples: six particles at fixed relative azimuths and a random overall orientation. This adds the known nonflow correlation. However, in the flow sample, for these reserved groups, the six azimuths are drawn independently for each cluster, so there is no nonflow correlation. The remaining tracks in the event, forming the bulk, are emitted from one common plane in the flow and mixed samples, with \(v_1=0\) and fixed \(v_2,\ldots,v_6\). In the nonflow sample, the bulk is drawn uniformly in azimuth, i.e., with no flow modulation. To remove any correlation between the nonflow cluster and the bulk, each cluster is rotated at random and is not aligned with the common plane, so particles from a cluster are uncorrelated with particles from the plane. The correlator measured in the mixed sample is then the sum of the correlator of the flow sample and the correlator of the nonflow sample. In the mixed sample, the \(70\)--\(100\%\) class is filled differently from the other eight classes. The clusters are generated as in the other classes, while the remaining tracks have uniform azimuths and are not attached to a common plane. The correlator of this class therefore contains the clusters and no flow. Equations~\eqref{eq:c0} and~\eqref{eq:c1} use it as \(C_{\mathrm{LM}}\). For the sake of simplicity, the values of \(v_2,\ldots,v_6\) are made multiplicity independent, and the relative azimuths of the tracks inside a cluster are given in Appendix~\ref{app:toy}.

\begin{figure}[!ht]
\centering
\includegraphics[width=0.9\linewidth]{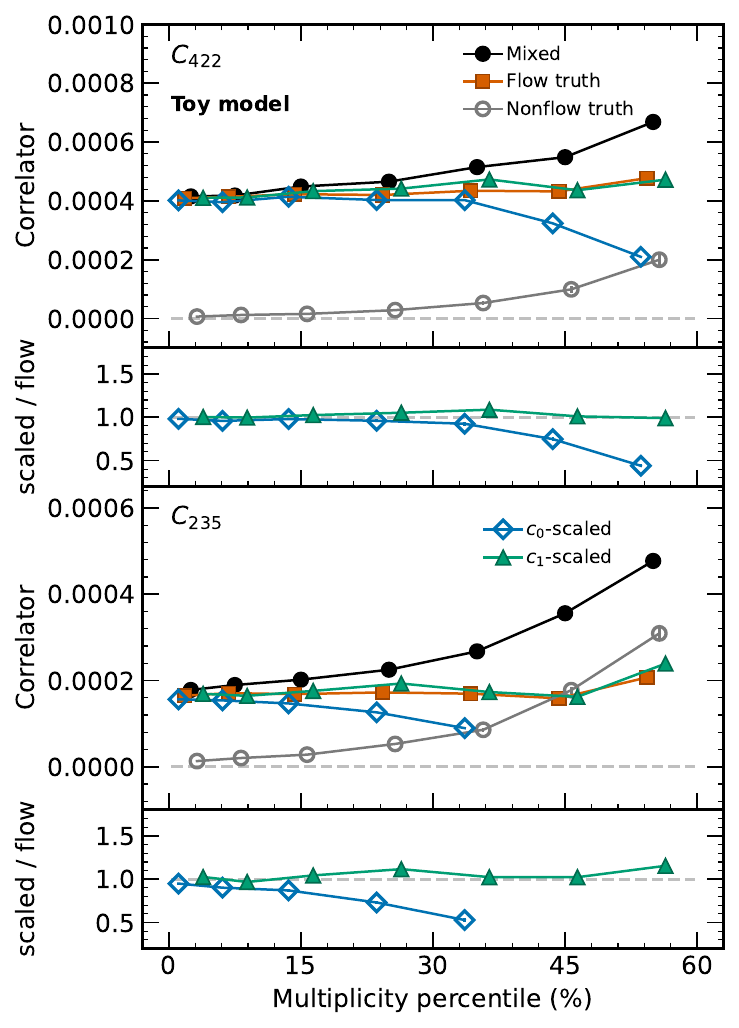}
\caption{Toy-model closure examples for the numerators \(C_{422}\) (top) and \(C_{235}\) (bottom).
Upper panels: mixed sample, flow truth, nonflow truth, and the mixed sample after \(\czero\) and \(\cone\) subtraction. Lower panels: subtracted correlator divided by the flow truth.}
\label{fig:toyC}
\end{figure}

Each sample is passed through the same recursive SPC analysis framework independently.
Equations~\eqref{eq:c0} and~\eqref{eq:c1} are applied only to the mixed sample, using its own LM bin. The flow and nonflow samples are the truth against which that result is compared. Figure~\ref{fig:toyC} shows the closure for the numerators \(C_{422}\) and \(C_{235}\) in the \(0\)--\(60\%\) multiplicity classes. In the upper panels, the mixed points follow an approximate sum of correlations from the flow truth and the nonflow truth. The gradual increase of correlations towards low-multiplicity events in the mixed sample arises purely from nonflow in this toy model. After \(\czero\) and \(\cone\) scalings are applied to the mixed sample, the nonflow-subtracted correlations move closer to the flow truth. The lower panel is the ratio of each subtracted correlator to the flow truth, which shows that the $c_1$ scaling works better than the $c_0$ scaling, as the latter seems to oversubtract nonflow towards the peripheral bins. The nonflow sample shows the contribution that the subtraction should remove.



\subsection{Pythia~8 as a nonflow reference}
\label{subsec:pythia}

To evaluate the performance of the proposed nonflow subtraction methods, the \textsc{Pythia~8 Angantyr} model~\cite{Sjostrand:2014zea,Bierlich:2022pfr,Bierlich:2018xfw} is used to generate approximately 165 million minimum-bias OO collisions at \(\sqrt{s_{\rm NN}}=5.36\)~TeV.
The nucleon--nucleon cross section uses the built-in parameters \((2.15,\,18.42,\,0.33)\), with no refit~\cite{Bierlich:2018xfw}.
Particles with \(c\tau>10\)~mm are left undecayed, following the ALICE primary-particle selection~\cite{ALICE:2017hcy}.
In this configuration, there are no collective flow components in the \textsc{Pythia~8 Angantyr} model. The azimuthal correlations come from sources such as jet production and fragmentation, multiparton interactions, resonance decays, and momentum conservation, which make the sample a nonflow-only reference.

\begin{figure*}[!ht]
\centering
\includegraphics[width=\linewidth]{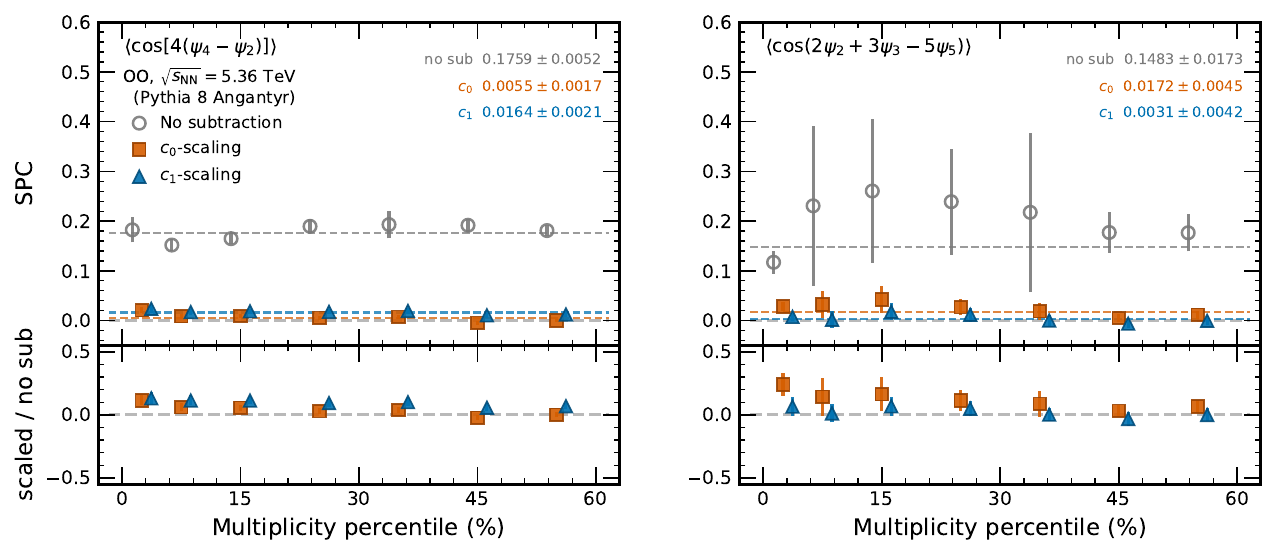}\\[0.4em]
{\footnotesize (a)}\\[0.6em]
\includegraphics[width=\linewidth]{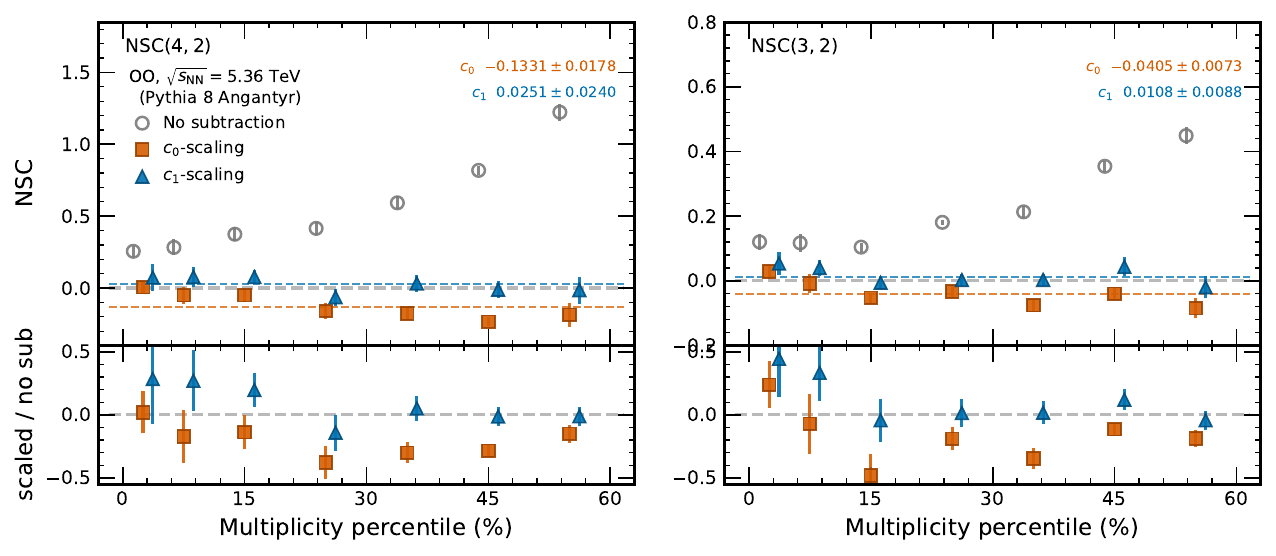}\\[0.4em]
{\footnotesize (b)}
\caption{Multiplicity-percentile dependence of multiparticle correlators in OO collisions at $\sqrt{s_{\mathrm{NN}}}=5.36$~TeV from \textsc{Pythia~8 Angantyr} showing the unsubtracted, $\czero$, and $\cone$-scaled correlations.
(a)~SPCs $\langle\cos[4(\psi_{4}-\psi_{2})]\rangle$ and $\langle\cos(2\psi_{2}+3\psi_{3}-5\psi_{5})\rangle$.
(b)~NSC$(4,2)$ and NSC$(3,2)$.
Dashed lines show constant fits where applicable.
Ratio panels show the scaled result divided by the unsubtracted one.}
\label{fig:pythiaSPC}
\end{figure*}

\begin{table}[t]
\caption{\label{tab:pythiaResiduals}
Residual nonflow in \textsc{Pythia~8 Angantyr} OO collisions at $\sqrt{s_{\mathrm{NN}}}=5.36$~TeV
after $\czero$ and $\cone$ subtraction.
Each entry is $100\times a_{c_{x}}/a_{\mathrm{nosub}}$, where $a$ is a constant (pol0) fit
to the multiplicity-percentile dependence for $0\text{--}60\%$ (residual fraction as in Ref.~\cite{Wang:2026hak}).
Uncertainties are from the fits.
A residual consistent with zero indicates successful nonflow removal in this flow-free sample.}
\begin{ruledtabular}
\begin{tabular}{lcc}
\textrm{Correlator} & $c_{0}$ residual [\%] & $c_{1}$ residual [\%] \\
\colrule
\multicolumn{3}{l}{Symmetry plane correlators (SPC)} \\
$\langle\cos[4(\psi_{4}-\psi_{2})]\rangle$ & $+3.1\pm1.0$ & $+9.3\pm1.2$ \\
$\langle\cos(2\psi_{2}+3\psi_{3}-5\psi_{5})\rangle$ & $+11.6\pm3.3$ & $+2.1\pm2.8$ \\
$\langle\cos[6(\psi_{6}-\psi_{2})]\rangle$ & $-20.7\pm15.8$ & $+39.1\pm22.4$ \\
$\langle\cos[6(\psi_{6}-\psi_{3})]\rangle$ & $+8.7\pm13.6$ & $+4.4\pm12.6$ \\
$\langle\cos(2\psi_{2}+4\psi_{4}-6\psi_{6})\rangle$ & $+3.9\pm4.2$ & $-3.0\pm4.2$ \\
\colrule
\multicolumn{3}{l}{Normalized symmetric cumulants (NSC)} \\
NSC$(4,2)$ & $-23.7\pm3.2$ & $+4.5\pm4.3$ \\
NSC$(3,2)$ & $-20.3\pm3.7$ & $+5.4\pm4.4$ \\
NSC$(4,3)$ & $+6.7\pm8.3$ & $+19.9\pm10.4$ \\
NSC$(5,2)$ & $+2.2\pm6.5$ & $+27.1\pm7.0$ \\
NSC$(5,3)$ & $-11.4\pm15.7$ & $+4.2\pm14.9$ \\
\end{tabular}
\end{ruledtabular}

\end{table}

In \textsc{Pythia} \(pp\) events, the standard multiparticle cumulants are dominated by dijets~\cite{Jia:2017hbm}. This has been used to study nonflow in symmetric cumulants~\cite{Huo:2017nms}, to test two-particle factorization and low-multiplicity template methods~\cite{Kikola:2011tu,Lim:2019cys,Nagle:2021rep,Ji:2023eqn}, and to study the correlation of \(v_2\) with the mean transverse momentum~\cite{Zhang:2021phk}. A multiplicity-scaled subtraction of multiparticle cumulants has been tested with \textsc{hijing} in OO collisions at the same energy~\cite{Wang:2026hak}. In this study, the \(\czero\) and \(\cone\) subtractions are applied to the SPCs and NSCs computed from the \textsc{Pythia} events. In the absence of collective flow, a complete nonflow subtraction should produce values consistent with zero for the SPCs and NSCs, and any remaining signal is a residual bias of the method.

Figure~\ref{fig:pythiaSPC} shows the multiplicity-percentile dependence of $\langle\cos[4(\psi_{4}-\psi_{2})]\rangle$,
$\langle\cos(2\psi_{2}+3\psi_{3}-5\psi_{5})\rangle$,
NSC$(4,2)$, and NSC$(3,2)$
in OO collisions at $\sqrt{s_{\mathrm{NN}}}=5.36$~TeV from \textsc{Pythia~8 Angantyr},
without and with $\czero$- and $\cone$-scaling-based nonflow subtractions. Even without collective flow, \textsc{Pythia~8 Angantyr} produces finite correlations for both SPCs and NSCs.
The NSCs grow towards low multiplicity, whereas the SPCs show little multiplicity dependence. This is the nonflow reference the methods need to subtract. When the nonflow subtraction is applied, both scalings drive the SPCs close to zero within uncertainties. For the NSCs, $\cone$ scaling performs better, giving values consistent with zero, while the $\czero$ scaling oversubtracts and leaves a negative residual.

To quantify the remaining nonflow bias, Table~\ref{tab:pythiaResiduals} reports the residual as the ratio of constant (pol0) fits to the subtracted and unsubtracted observables over the $0\text{--}60\%$ multiplicity-percentile range.
Both methods remove most of the nonflow, and for most entries the residual is at the $10\%$ level or below.
For the SPCs, the two methods give similar results, with $\cone$ providing slightly better overall nonflow control.
One exception is the four-particle SPC $\langle\cos[6(\psi_{6}-\psi_{2})]\rangle$, for which neither method works well: $\czero$ and $\cone$ leave residuals of $-20\%$ and $+39\%$, respectively.
For the NSCs, the difference between the methods is more pronounced.
For NSC$(4,2)$, NSC$(3,2)$, and NSC$(5,3)$, the $\cone$ residual is consistent with zero, while $\czero$ oversubtracts by about $20\%$.
For NSC$(4,3)$ and NSC$(5,2)$, $\czero$ performs better, whereas $\cone$ leaves a positive residual of around $20\%$. For completeness, the remaining SPC and NSC observables are collected in Appendix~\ref{app:pythiaRest}.

It is important to note that, for the SCs and NSCs, nonflow can also be suppressed using the subevent method, where the particles of the correlator are taken from two or more separate pseudorapidity regions (subevents), with or without a pseudorapidity gap between them~\cite{Jia:2017hbm,Huo:2017nms}.
We have checked this approach in the present setup and found that it reduces the statistical precision, which gets worse with an increase in the number of subevents and in the size of the pseudorapidity gap.
Thus, the subevent methods are not included in the present study.


\begin{figure*}[!ht]
\centering
\includegraphics[width=\linewidth]{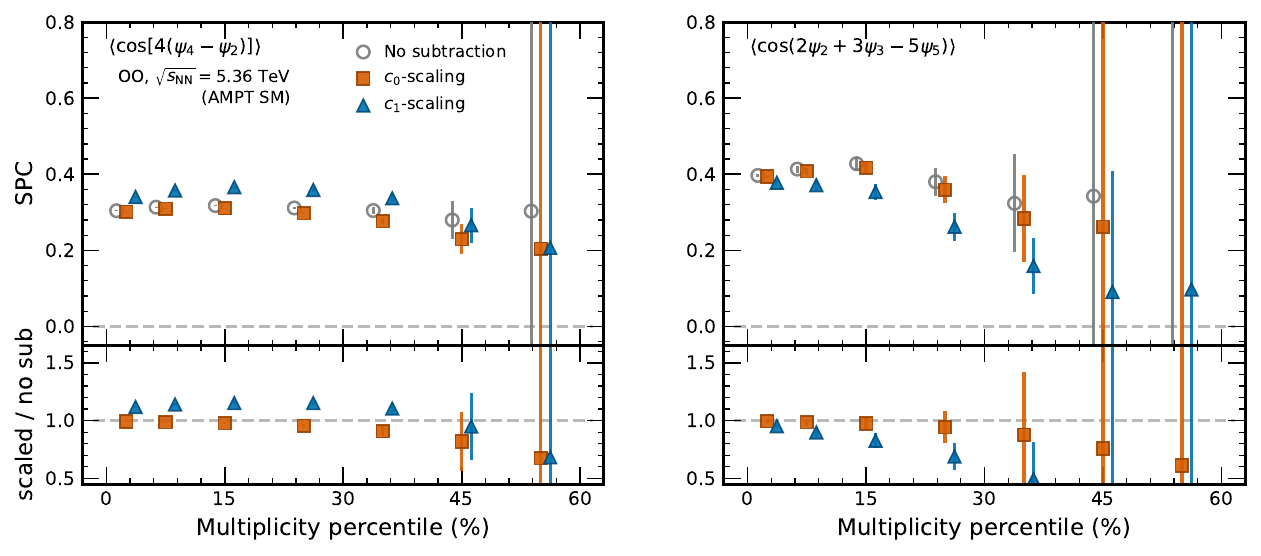}\\[0.4em]
{\footnotesize (a)}\\[0.6em]
\includegraphics[width=\linewidth]{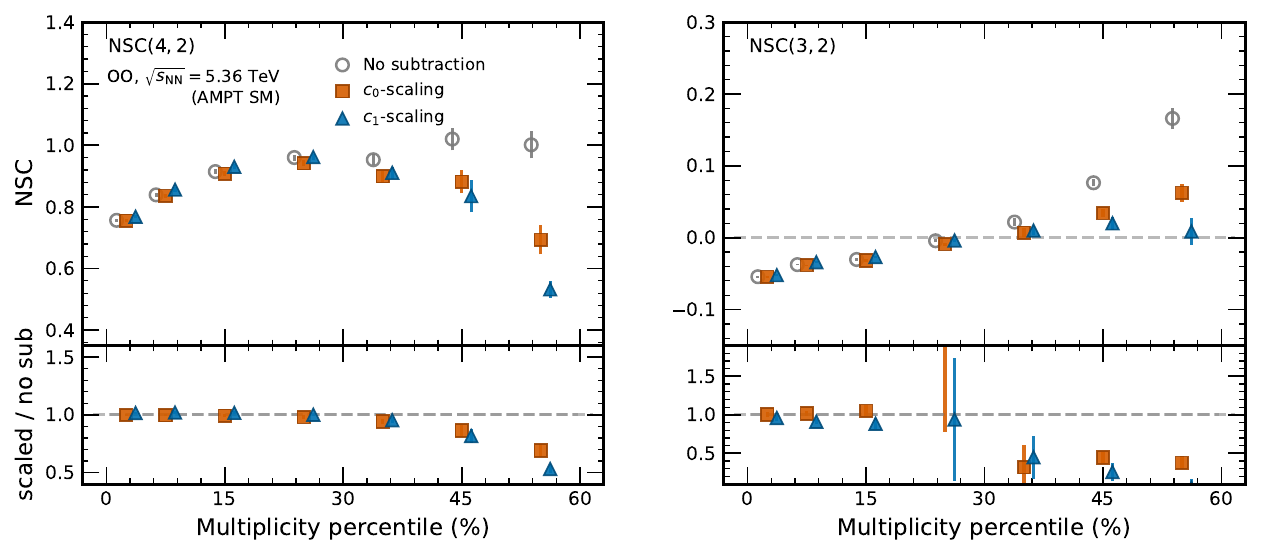}\\[0.4em]
{\footnotesize (b)}
\caption{Multiplicity-percentile dependence of multiparticle correlators in OO collisions at $\sqrt{s_{\mathrm{NN}}}=5.36$~TeV from the \textsc{ampt} model showing the unsubtracted, $\czero$, and $\cone$-scaled correlations.
(a)~SPCs \(\langle\cos[4(\psi_{4}-\psi_{2})]\rangle\) and \(\langle\cos(2\psi_{2}+3\psi_{3}-5\psi_{5})\rangle\).
(b)~NSC$(4,2)$ and NSC$(3,2)$.
Ratio panels show the scaled result divided by the unsubtracted one.}
\label{fig:amptSPC}
\end{figure*}

\section{Results and discussion}
\label{sec:results}

The \textsc{ampt} model provides a realistic testing ground for nonflow-control methods, as its final-state correlations reflect the interplay of collective flow and nonflow effects (see Appendix~\ref{app:ampt} for a brief model
description). The collective-flow signals emerge primarily from the response of the partonic medium to initial-state geometry and fluctuations~\cite{Lin:2004en}, while nonflow contributions can arise from minijets, fragmentation, and resonance decays, which can be smeared and modified by hadronization and hadronic rescattering~\cite{Ma:2010dv, Xu:2012ue, Jia:2013tja, Nie:2018xog}. Although it is not trivial to uniquely disentangle the flow and nonflow components in the model, their coexistence allows the robustness and limitations of the proposed nonflow subtraction methods to be systematically checked.

\begin{figure*}[!ht]
\centering
\includegraphics[width=\linewidth]{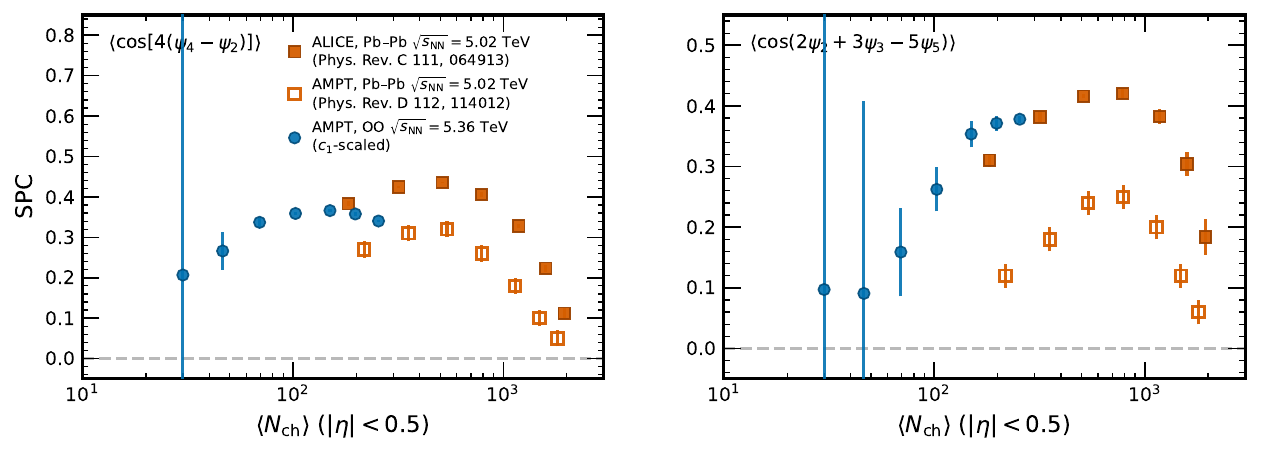}\\[0.4em]
{\footnotesize (a)}\\[0.6em]
\includegraphics[width=\linewidth]{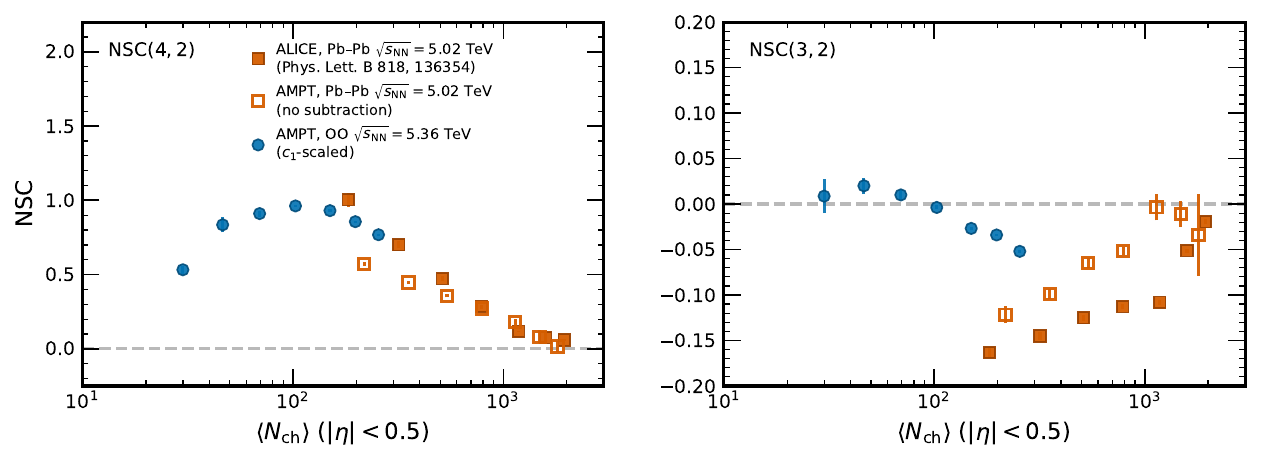}\\[0.4em]
{\footnotesize (b)}
\caption{Nonflow-subtracted ($c_1$-scaled) multiparticle correlators measured as a function of charged-particle multiplicity at midrapidity in OO collisions at $\sqrt{s_{\mathrm{NN}}}=5.36$~TeV from the \textsc{ampt} model. Results are compared with Pb--Pb collisions at $\sqrt{s_{\mathrm{NN}}}=5.02$~TeV from ALICE~\cite{ALICE:2024fus,ALICE:2021adw} and \textsc{ampt}~\cite{Tripathy:2025npe}.
(a)~SPCs \(\langle\cos[4(\psi_{4}-\psi_{2})]\rangle\) and \(\langle\cos(2\psi_{2}+3\psi_{3}-5\psi_{5})\rangle\).
(b)~NSC$(4,2)$ and NSC$(3,2)$.}
\label{fig:ooVsPbPb}
\end{figure*}

Figure~\ref{fig:amptSPC} shows the multiplicity-percentile dependence of
$\langle\cos[4(\psi_{4}-\psi_{2})]\rangle$,
$\langle\cos(2\psi_{2}+3\psi_{3}-5\psi_{5})\rangle$,
NSC$(4,2)$, and NSC$(3,2)$
in OO collisions at $\sqrt{s_{\mathrm{NN}}}=5.36$~TeV from the \textsc{ampt} string-melting model~\cite{Lin:2004en},
before and after $\czero$ and $\cone$ subtraction.
The model produces finite, multiplicity-dependent correlations for both SPCs and NSCs.
For multiplicity percentiles $\gtrsim 40\%$, the correlators rise abruptly towards more peripheral classes,
indicating growing nonflow contamination at low multiplicity.
Both $\czero$ and $\cone$ scalings remove this peripheral enhancement and restore a smoother
multiplicity dependence in the low-multiplicity region.
In more central (high-multiplicity) collisions the subtraction changes the values only mildly:
the scaled points remain close to the unsubtracted \textsc{ampt} results,
consistent with a weaker relative nonflow contribution where the multiplicity is large.

Motivated by the toy-model closure and the \textsc{Pythia~8} null test, where $\cone$ provides the more reliable nonflow control---particularly for NSC$(4,2)$ and NSC$(3,2)$---and by the \textsc{ampt} results, where $\czero$ and $\cone$ perform similarly with $\cone$ slightly preferred, we recommend $\cone$ scaling as the best choice for studying $\langle\cos[4(\psi_{4}-\psi_{2})]\rangle$, $\langle\cos(2\psi_{2}+3\psi_{3}-5\psi_{5})\rangle$, NSC$(4,2)$, and NSC$(3,2)$ in OO collisions. Figure~\ref{fig:ooVsPbPb} compares the $\cone$-scaled SPCs and NSCs in OO collisions at $\sqrt{s_{\rm NN}}=5.36$~TeV from \textsc{ampt} with Pb--Pb collisions at $\sqrt{s_{\rm NN}}=5.02$~TeV from ALICE~\cite{ALICE:2024fus,ALICE:2021adw} and also from \textsc{ampt}~\cite{Tripathy:2025npe}. To place both collision systems on an equal footing, the multiplicity-percentile bins are mapped to the corresponding mean charged-particle multiplicity at midrapidity, $\langle N_{\rm ch}\rangle(|\eta|~<~0.5)$~\cite{ALICE:2015juo}. This comparison is preferable because the relation between collision centrality and impact parameter---and thus the connection to the collision geometry---is weaker in light-ion collisions than in heavy-ion collisions. Plotting versus $\langle N_{\rm ch}\rangle$ furthermore allows the system-size dependence of these correlators to be examined across OO and Pb--Pb in a common variable~\cite{ALICE:2019zfl}.

In Fig.~\ref{fig:ooVsPbPb}(a), both $\langle\cos[4(\psi_{4}-\psi_{2})]\rangle$ and $\langle\cos(2\psi_{2}+3\psi_{3}-5\psi_{5})\rangle$ in \textsc{ampt} OO increase from low- to high-multiplicity events. The correlator, $\langle\cos[4(\psi_{4}-\psi_{2})]\rangle$, rises to a broad maximum near $\langle N_{\rm ch}\rangle\sim150$ and then decreases mildly towards the most central OO bin, whereas the three-harmonic correlator, $\langle\cos(2\psi_{2}+3\psi_{3}-5\psi_{5})\rangle$, continues to increase up to the highest OO multiplicity ($\langle N_{\rm ch}\rangle\sim250$). In the overlap region around $\langle N_{\rm ch}\rangle\sim200$--$250$, the OO values reach a magnitude comparable to peripheral Pb--Pb. Extending into the Pb--Pb domain, both correlators attain their global maxima in mid-central collisions (roughly $\langle N_{\rm ch}\rangle\sim500$--$800$) and then decrease towards the most central, highest-multiplicity Pb--Pb events. \textsc{ampt} underpredicts the Pb--Pb correlators relative to ALICE, while reproducing the qualitative multiplicity dependence. This suggests that OO data may follow a similar trend to \textsc{ampt} OO, although the absolute magnitudes could differ.

For NSC$(4,2)$, the correlator in \textsc{ampt} OO rises steeply with multiplicity, peaks near $\langle N_{\rm ch}\rangle\sim100$, and already declines towards the highest OO multiplicities. Across the OO--Pb--Pb overlap it connects smoothly in magnitude to peripheral Pb--Pb, after which the Pb--Pb values fall steadily towards zero in the most central collisions. \textsc{ampt} describes the Pb--Pb data well at high multiplicity and underpredicts them somewhat in more peripheral (lower-multiplicity) Pb--Pb bins. For NSC$(3,2)$, the OO correlator remains close to zero from low to mid multiplicity and becomes only modestly negative ($\sim-0.05$) in the highest-multiplicity OO bin. In Pb--Pb, by contrast, the correlator starts more strongly negative ($\sim-0.16$) in peripheral collisions and approaches zero towards the most central events. \textsc{ampt} again reproduces the qualitative Pb--Pb trend but underpredicts the magnitude of the anticorrelation. Therefore, the NSCs in OO data may also follow a similar trend to \textsc{ampt} OO, although the magnitudes could differ.

The Pb--Pb trends are driven by how initial geometry and hydrodynamic response change with system size. ALICE finds that $\langle\cos[4(\psi_{4}-\psi_{2})]\rangle$ and related plane correlators are weakest in central collisions, where the overlap is nearly isotropic, and strongest in mid-central collisions, where the almond-shaped geometry and the nonlinear medium response that couples $\psi_{4}$ to $\psi_{2}$ (and $\psi_{5}$ to $\psi_{2}$ and $\psi_{3}$) are most pronounced, before easing again towards the periphery~\cite{ALICE:2024fus}. Likewise, NSC$(4,2)$ is positive because event-by-event $v_{2}$ and $v_{4}$ fluctuations are correlated through both the initial $\varepsilon_{2}$--$\varepsilon_{4}$ geometry and the nonlinear $v_{4}\leftarrow\varepsilon_{2}$ response (hence NSC$(4,2)$ is also sensitive to $\eta/s$), while NSC$(3,2)$ is negative because $v_{2}$ and $v_{3}$ inherit the anticorrelation of $\varepsilon_{2}$ and $\varepsilon_{3}$, making NSC$(3,2)$ largely an initial-state probe~\cite{ALICE:2016kpq,ALICE:2017kwu,ALICE:2021adw}. Mapped versus $\langle N_{\rm ch}\rangle$, the \textsc{ampt} OO results sit on the rising, smaller-system side of this picture: as multiplicity grows within OO, the overlap develops a clearer anisotropic geometry and nonlinear couplings strengthen, so the SPCs and NSC$(4,2)$ increase towards the peripheral Pb--Pb values near $\langle N_{\rm ch}\rangle\sim200$, whereas NSC$(3,2)$ remains near zero until the system is large enough for a Pb--Pb-like $\varepsilon_{2}$--$\varepsilon_{3}$ anticorrelation to emerge. Continuing into Pb--Pb, the correlators then follow the familiar mid-central maximum (SPCs) or the fall of NSC$(4,2)$ towards zero and the approach of NSC$(3,2)$ towards zero in the most central, nearly isotropic collisions.

\section{Summary and outlook}
\label{sec:summary}

The multiplicity-reweighted ($\czero$) and $v_1$-based-estimator
($\cone$) nonflow subtraction methods are studied for SPCs and NSCs in OO collisions at
$\sqrt{s_{\rm NN}}=5.36$~TeV, and the preferred prescription is used to predict these correlators in the \textsc{ampt} model and to compare them with results from Pb--Pb collisions. The important findings are summarized below.

\begin{itemize}
  \item In a toy model with known flow and nonflow components, both scalings remove the nonflow rise towards low multiplicity; $\cone$ tracks the flow truth more closely, while $\czero$ tends to oversubtract the collective-flow-like correlations in peripheral bins.
  
  \item In the \textsc{Pythia~8 Angantyr} model, used as a nonflow-only reference, both methods drive most SPCs close to zero (residuals typically $\lesssim 10\%$). For NSC$(4,2)$ and NSC$(3,2)$, $\cone$ leaves residuals consistent with zero, whereas $\czero$ oversubtracts by about $20\%$.
  
  \item In the \textsc{ampt} model, $\czero$ and $\cone$ both remove the peripheral nonflow enhancement and leave the high-multiplicity points nearly unchanged in OO collisions; following the closure tests, the $\cone$ scaling is recommended for the main SPCs and NSCs studied in this work.
  
  \item When studied as a function of $\langle N_{\rm ch}\rangle$, the $\cone$-scaled $\langle\cos[4(\psi_{4}-\psi_{2})]\rangle$, $\langle\cos(2\psi_{2}+3\psi_{3}-5\psi_{5})\rangle$, and NSC$(4,2)$ in OO collisions from \textsc{ampt} rise with multiplicity and approach peripheral Pb--Pb collisions near $\langle N_{\rm ch}\rangle\sim 200$. On the other hand, NSC$(3,2)$ in OO collisions stays near zero and only attains a modest negative value towards high multiplicity. Results from Pb--Pb collisions then show the familiar mid-central SPC maximum and the decline (approaching zero) of NSC$(4,2)$ [NSC$(3,2)$] towards central collisions, consistent with evolving initial geometry and nonlinear hydrodynamic response.
  
\end{itemize}

These predictions from the \textsc{ampt} model, after $\cone$ subtraction and versus $\langle N_{\rm ch}\rangle$, provide a concrete baseline for forthcoming SPC and NSC measurements in $p$O, OO, and Ne--Ne collisions at the LHC. Comparing data to this baseline and extending the same nonflow framework to additional correlators and nuclear geometries will provide necessary constraints on models incorporating initial-state effects, transport properties, and medium evolution, and on the onset of nonlinear mode coupling from small to large systems.

\begin{acknowledgments}
The authors would like to thank Dr. Dong Jo Kim, University of Jyv\"askyl\"a, Finland, for fruitful discussions. NM is supported by the Academy of Finland through the Center of Excellence in Quark Matter (Grant No. 346328). SP gratefully acknowledges support from the Hungarian National Research, Development and Innovation Office (NKFIH) under Contract Nos. NKFIH NKKP ADVANCED\_25 K153456, 2025-1.1.5-NEMZ\_KI-2025-00005, 2024-1.2.5-TET-2024-00022, FuSe COST action CA241011. The authors also acknowledge the Wigner Scientific Computing Laboratory (WSCLAB) in Budapest, Hungary for allocating computational resources for this study.

\end{acknowledgments}

\appendix

\section{Parameters of the toy model}
\label{app:toy}

The analysis uses the same nine multiplicity percentile bins as the main analysis, and the corresponding multiplicity bin edges are,
\[
0,\; 25,\; 40,\; 58,\; 83,\; 113,\; 152,\; 202,\; 238,\; 3000,
\] ordered from low to high multiplicity.  The interval from \(0\) to \(25\) is the most peripheral class, 70--100\%, and the interval from \(238\) to \(3000\) (the last edge is an upper bound, not a typical multiplicity) is the most central class, 0--5\%. The event multiplicity ($N$) is drawn randomly from a Gaussian distribution with mean $\langle N \rangle$ and standard deviation $0.12\langle N \rangle$ for each event class. From central to peripheral bins, the nine means are
\[
\langle N\rangle = 270,\; 220,\; 175,\; 130,\; 98,\; 68,\; 48,\; 32,\; 24.
\] 
Draws that fall outside the class, or below \(N=18\), are rejected.
The requirement \(N\ge 18\) leaves room for one group of six tracks (for the nonflow cluster) and twelve tracks for the bulk formation which carry the flow modulation.

The number of groups ($N_s$) is found with 
\begin{equation}
N_s=\max\bigl(1,\lfloor f N/6\rfloor\bigr),
\qquad f=0.18,
\label{eq:nsblocks}
\end{equation}
and \(N_s\) is reduced if the bulk would otherwise contain fewer than \(12\) tracks. For example, if the event has 100 particles, $N_s = 3$, so the reserved clusters are made with $3\times6=18$ tracks, and 82 tracks are left for the bulk.

A particle emitted from the common plane with plane angle $\psi$ is drawn from
\begin{equation}
\frac{\mathrm{d}N}{\mathrm{d}\varphi}
\propto
1+2\sum_{n=2}^{6} v_n\cos\bigl[n(\varphi-\psi)\bigr],
\label{eq:toyflow}
\end{equation}
where $\psi$ is chosen uniformly in $[0,2\pi)$, and $\varphi$ is the azimuthal angle of the particle. 
The flow coefficients are kept the same in every class,
\begin{gather*}
v_1=0,\\
(v_2,v_3,v_4,v_5,v_6)=(0.12,\,0.07,\,0.05,\,0.035,\,0.025).
\end{gather*}
For the nonflow clusters of six particles each, the six relative azimuths are
\[
\delta_k=(0,\,5.9072,\,3.3545,\,3.6753,\,0.0195,\,6.0169),
\]
in radians. Track \(k\) is placed at \(\varphi_k=\alpha+\delta_k\), with a new angle \(\alpha\), drawn uniformly, for every cluster. A track that belongs neither to the plane nor to a cluster is given an azimuth drawn uniformly on \([0,2\pi)\).
In the LM class, this uniform azimuth is used for the bulk of every sample (no flow is assumed in the LM class), while the mixed and nonflow samples still fill their groups with clusters. For every track, whatever its azimuth, \(\eta\) is uniform on \((-0.8,0.8)\), \(p_{\mathrm{T}}\) is uniform on \((0.2,3)\)~GeV/\(c\), and the charge is random. None of these depends on \(\varphi\), so the only azimuthal correlations are the common plane and the clusters.

\begin{figure}[!t]
\centering
\includegraphics[width=\linewidth]{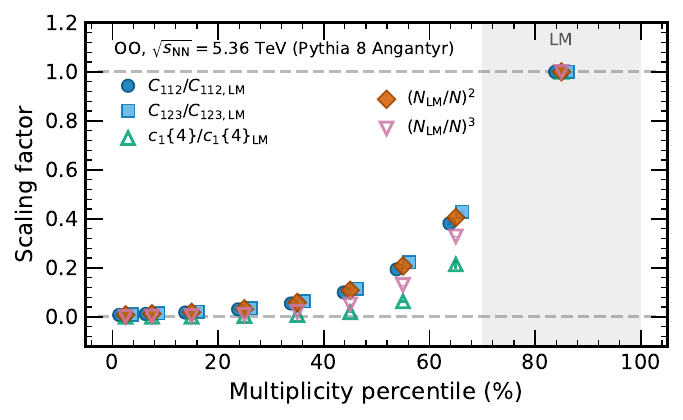}
\includegraphics[width=\linewidth]{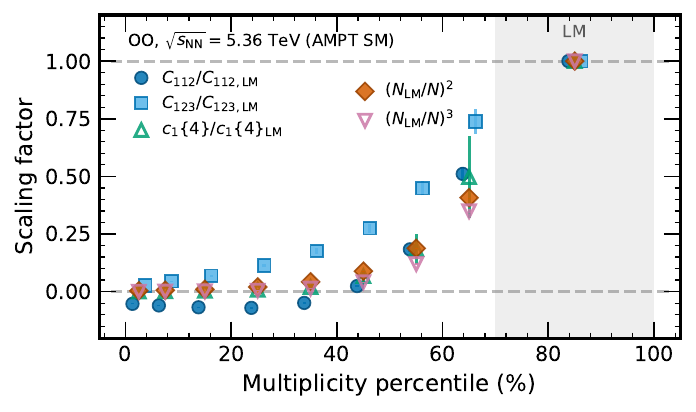}
\caption{Scaling factors used for the numerators in various SPCs versus multiplicity percentiles in OO collisions at $\sqrt{s_{\rm NN}} = 5.36$~TeV from \textsc{Pythia~8} (top) and the \textsc{ampt} model (bottom). The gray band marks the LM class.}
\label{fig:scaling}
\end{figure}

\section{AMPT model}
\label{app:ampt}
A Multi-Phase Transport (\textsc{ampt}) model is a Monte Carlo transport framework developed to describe the evolution of relativistic nuclear collisions through the initial-state, partonic, hadronization, and hadronic stages~\cite{Lin:2004en}. The initial conditions are generated using the Heavy Ion Jet Interaction Generator (\textsc{hijing})~\cite{Wang:1991hta}, while partonic scatterings are described by Zhang's Parton Cascade (ZPC) model~\cite{Zhang:1998zpc}. Hadronization proceeds through Lund string fragmentation in the default version of \textsc{ampt}, whereas, in the string-melting version, the excited strings are converted into partons that subsequently form hadrons through quark coalescence~\cite{Andersson:1983,Lin:2001zk,Lin:2004en}. The produced hadrons can undergo further interactions described by the A Relativistic Transport (ART) model~\cite{Li:1995pra}. Here, the \textsc{ampt} settings used for OO collisions are similar to those in Ref.~\cite{Behera:2021zhi}, while for Pb--Pb collisions, the same settings as in Ref.~\cite{Tripathy:2025npe} are used. This framework enables investigations of how initial-state fluctuations, partonic interactions, hadronization, and hadronic rescattering influence final-state particle production and azimuthal correlations. \textsc{ampt} has consequently been used to study anisotropic flow and multiparticle correlations, providing a means of investigating the interplay between initial-state geometry and final-state dynamics~\cite{Lin:2004en,Nie:2018xog,Nasim:2016rfv}. Studies based on \textsc{ampt} in OO collisions have investigated anisotropic flow, normalized symmetric cumulants, two-particle azimuthal correlations, and global observables for different oxygen nuclear configurations, including Woods--Saxon, harmonic-oscillator, and some exotic $\alpha$-clustered geometries~\cite{Behera:2023nwj,Behera:2021zhi,Zhao:2024feh,Prasad:2026osg}. These studies demonstrate the relevance of the \textsc{ampt} model for predicting collective-flow-related observables in light-ion collisions and motivate further investigations of correlations between different flow harmonics such as the SPCs and NSCs~\cite{ALICE:2016kpq,ALICE:2017kwu,Nasim:2016rfv}.

\begin{figure*}[!ht]
\centering
\includegraphics[width=0.95\linewidth]{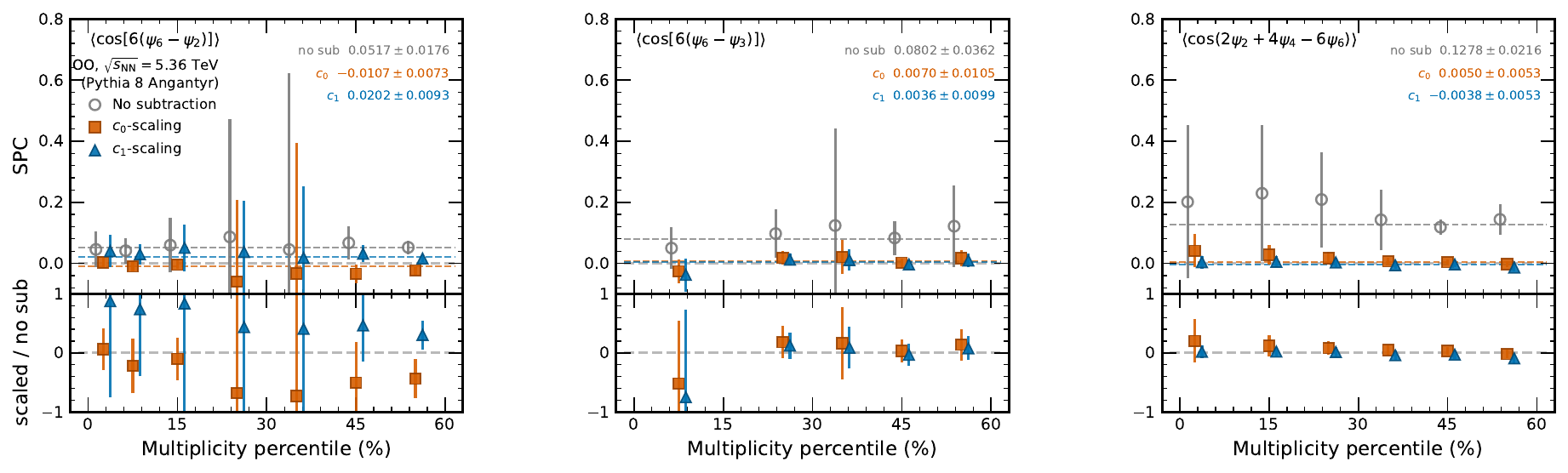}\\[0.4em]
{\footnotesize (a)}\\[0.6em]
\includegraphics[width=0.95\linewidth]{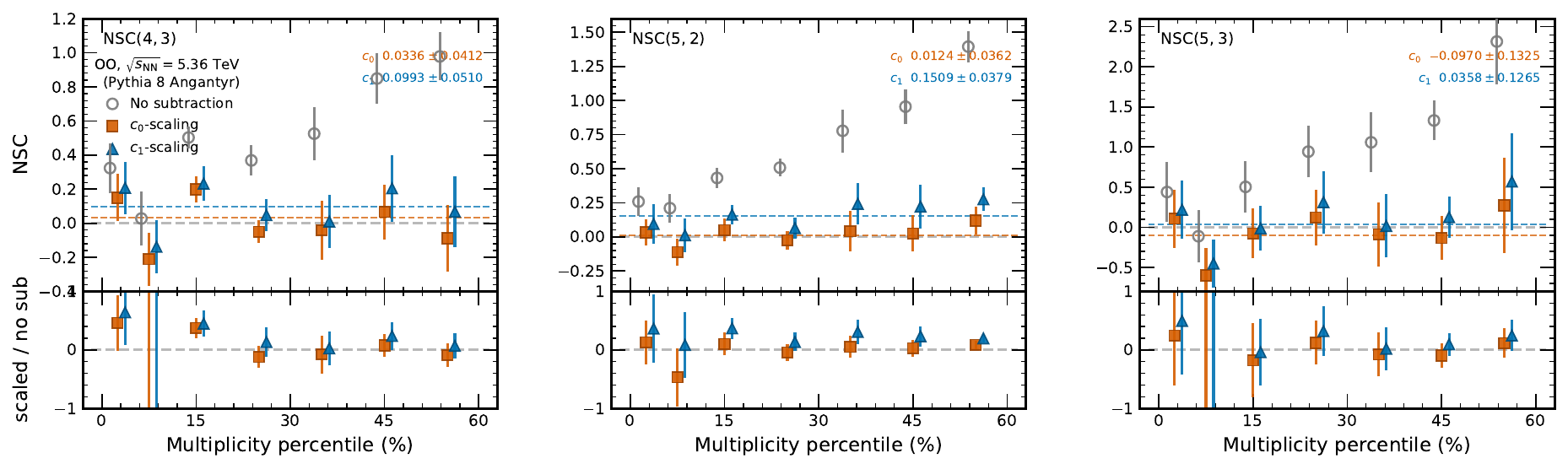}\\[0.4em]
{\footnotesize (b)}
\caption{Multiplicity percentile dependence of SPCs and NSCs in OO collisions at $\sqrt{s_{\mathrm{NN}}}=5.36$~TeV from \textsc{Pythia~8 Angantyr} model:
no subtraction, $\czero$, and $\cone$.
(a)~Additional SPCs
$\langle\cos[6(\psi_{6}-\psi_{2})]\rangle$,
$\langle\cos[6(\psi_{6}-\psi_{3})]\rangle$, and
$\langle\cos(2\psi_{2}+4\psi_{4}-6\psi_{6})\rangle$.
(b)~Additional NSCs: NSC$(4,3)$, NSC$(5,2)$, and NSC$(5,3)$.
Dashed lines show constant fits where applicable; ratio panels are
scaled/no-subtraction.}
\label{fig:pythiaRest}
\end{figure*}

\section{Scaling factors}

The nonflow subtractions of Eqs.~\eqref{eq:c0} and~\eqref{eq:c1}
are fixed by the scaling factors
\begin{equation}
S_{c_0}
=
\left(\frac{N_{m,\mathrm{LM}}}{N_m}\right)^{m-1},
\qquad
S_{c_1}
=
\frac{E}{E_{\mathrm{LM}}},
\end{equation}
where \(N_m\) is the effective multiplicity of Eq.~\eqref{eq:Neff}
and \(E\) is the nonflow estimator of the same particle order as the
target correlator (Table~\ref{tab:spcE}).
Both factors are defined to equal unity in the LM reference class
($70$--$100\%$) and to fall below one towards higher multiplicity,
where the relative nonflow weight is smaller.
Figure~\ref{fig:scaling} shows these scaling factors as a function of multiplicity percentiles in OO collisions at $\sqrt{s_{\rm NN}} = 5.36$~TeV from \textsc{Pythia~8} (top) and the \textsc{ampt} model (bottom).
For \(\cone\), the three-particle estimators
$C_{112}/C_{112,\mathrm{LM}}$ and $C_{123}/C_{123,\mathrm{LM}}$ are
used for the three-particle SPC numerators, and the four-particle
estimator $c_{1}\{4\}/c_{1}\{4\}_{\mathrm{LM}}$ is used for
$C_{6222}$.
For \(\czero\), $(N_{3,\mathrm{LM}}/N_3)^{2}$ and
$(N_{4,\mathrm{LM}}/N_4)^{3}$ correspond to three- and
four-particle correlators, respectively.
The gray band marks the LM class.
The \(\cone\) factors track the multiplicity trend of the
\(v_1\)-based estimators and can differ from the pure
multiplicity powers \(S_{c_0}\); that difference is what allows
\(\cone\) to absorb residual multiplicity-selection effects beyond
the independent-source $1/N^{m-1}$ baseline.

\section{Additional Pythia SPCs and NSCs}
\label{app:pythiaRest}
Figure~\ref{fig:pythiaRest} shows the remaining SPCs and NSCs in
\textsc{Pythia~8 Angantyr} OO collisions after $\czero$ and $\cone$
subtraction.
As for the main-text observables, a successful subtraction should drive
the correlators to zero in this flow-free sample.
Both scalings bring
$\langle\cos[6(\psi_{6}-\psi_{3})]\rangle$ and
$\langle\cos(2\psi_{2}+4\psi_{4}-6\psi_{6})\rangle$
close to zero, whereas
$\langle\cos[6(\psi_{6}-\psi_{2})]\rangle$
retains a sizeable residual with either method.
For the NSCs, $\cone$ performs well for NSC$(5,3)$, while $\czero$ is
preferred for NSC$(4,3)$ and NSC$(5,2)$; the corresponding residual
fractions are listed in Table~\ref{tab:pythiaResiduals}.

\bibliography{refs}

\end{document}